\documentclass[aps,onecolumn,showpacs,floatfix,nofootinbib,superscriptaddress,11pt]{revtex4-1}
\usepackage{amssymb,amsmath}
\usepackage{graphicx,float,color}
\usepackage{indentfirst}
\usepackage{mathtools,harmony,upgreek}
\usepackage{color,accents,stmaryrd}
\usepackage{mathrsfs}
\usepackage{amssymb}
\usepackage{amsmath}
\usepackage{dsfont}
\usepackage{indentfirst}
\usepackage{latexsym}
\newcommand{\cl}{C \kern -0.1em \ell}

\newcommand{\p}{\partial}

\newcommand{\D}{\Delta}

\newcommand{\m}{\mu}
\newcommand{\n}{\nu}

\newcommand{\s}{\sigma}

\newcommand{\g}{\gamma}

\newcommand{\be}{\begin{equation}}
\newcommand{\ee}{\end{equation}}
\newcommand{\ba}{\begin{array}}
\newcommand{\ea}{\end{array}}
\newcommand{\f}{\frac}

\newcommand{\beq}{\begin{eqnarray}}
\newcommand{\eeq}{\end{eqnarray}}

\newcommand{\e}{\epsilon}
\newcommand{\w}{\wedge}

\newcommand{\om}{\Omega}

\renewcommand{\b}{\beta}

\renewcommand{\d}{\delta}
 
\renewcommand{\a}{\alpha}

\newcommand{\Db}{\bullet}%\mathring{\diamond}\;}

\newcommand{\vp}{\varphi}

\definecolor{clearblue}{rgb}{0,0.5,0.9}
\definecolor{orange}{rgb}{1,0.5,0}

\begin{document}

\title{New fermions in the bulk}

\author{K. P. S. de Brito}
\email{kelvyn.paterson@ufabc.edu.br}
 \affiliation{CCNH,  
Universidade Federal do ABC,\\ 09210-580, Santo Andr\'e, Brazil}

\author{Rold\~ao~da~Rocha}
\email{roldao.rocha@ufabc.edu.br}
\affiliation{CMCC, 
Universidade Federal do ABC,\\ 09210-580, Santo Andr\'e, Brazil.}


\begin{abstract}
Spinor fields on 5-dimensional Lorentzian manifolds are classified, according to the geometric Fierz identities that involve their  bilinear covariants. 
Based upon this classification that generalises the celebrated 4-dimensional Lounesto classification of spinor fields, new non-trivial classes of 5-dimensional spinor fields are, hence, found, with important potential applications regarding bulk fermions and their subsequent localisation on brane-worlds. In addition, quaternionic bilinear covariants are 
used to derive the quaternionic spin density, through the truncated exterior bundle. In order to accomplish a realisation of these new spinors, a Killing vector field is constructed on the horizon of 5-dimensional Kerr black holes. This Killing vector field  is shown to  reach the time-like Killing vector field at the spatial infinity, through a current 1-form density, constructed with the derived new spinor fields. The current density is, moreover, expressed as  the \emph{f\"unfbein} components, assuming a condensed form.
\end{abstract}
\maketitle
%\flushbottom

\section{Introduction}

Brane-world models represent an attempt to settle a variety of open questions in physics. These models must present a consistent effective 4-dimensional (4D) physics describing our Universe, for appropriate  limits \cite{Arkani02}. Among fields that should be localised on the brane, fermionic fields play a very important role to represent matter of the Standard Model \cite{DeWolfe}, whose zero modes, at least, are  localised \cite{Zhang,Andrianov:2013vqa,Liu0907.0910}.  Several alternative scenarios have been further studied to localise fermions 
on 3-branes generated by different models \cite{Bazeia:2004dh,German:2012rv,Bernardini:2014vba,Ahmed,Cruz:2013zka,casana}. 
The limitation regarding these models, despite of their laudable efforts to localise fermions, consists of the assumption, almost on their totality, of the 5-dimensional (5D) Dirac spinors paradigm.   Fermionic matter, in a 5D thick brane setup, is strictly based upon solving the Dirac equation in the bulk \cite{dzhu}. The sole exception  consist of the well-know mass dimension one eigenspinors of the charge conjugation operator with dual helicity \cite{lee1,daSilva:2012wp,exotic}, realised by the singular flagpole Elko spinors setup, on thick branes \cite{Jardim:2014xla,Liu:2011nb}. Since Dirac spinor fields represent a tiny class of regular spinors in the Lounesto spinor field classification, our main aim here is to show  that in a 5D bulk there are, similarly, new spinors classes, that can be further explored in physical theories, as well as their formal aspects.

The Lounesto spinor field classification in 4D Minkowski spacetime \cite{lou2}, according to spinor bilinear covariants, provides classes of regular and singular spinors.  In fact, recently all spinor fields have been thoroughly discriminated and scrutinised in successful classifications that are complementary to the Lounesto one \cite{Cavalcanti:2014wia,fabbri}, with many applications in field theory and gravitation \cite{daSilva:2012wp,esk,exotic,lee1}. 
The pivotal goal, that underlies the well-established Lounesto classification -- and further the 
more generalised classifications \cite{Cavalcanti:2014wia,fabbri}, is to find the  underlying spinors dynamics  in all classes, since just few spinors in the Lounesto classification have an identified dynamics associated to \cite{daSilva:2012wp}. In fact, besides the well-known Dirac, Weyl, and Majorana equations, recently flag-dipoles singular spinors -- that are type-4 spinors in Lounesto classification -- have been shown to be solutions of the Dirac equation with torsion, in an $f(R)$-gravity background, representing Einstein-Sciama-Kibble (ESK) models of gravity. Hence, singular flag-dipole spinor fields were used as a source of matter \cite{fabbri,esk}.
Moreover, 
flag-dipoles and flagpoles have been identified with singular solutions of the Dirac equation, in a special Kerr background solution \cite{Mei2,kerr}. Moreover, flagpoles singular spinors comprise realisations of 
mass dimension one (Elko) spinors, that are prime candidate for solving the dark matter problem  \cite{daSilva:2012wp,exotic,lee1}, whose Lagrangian has been very recently described  \cite{lee2}.  Out of the Dirac spinors paradigm,  singular spinors have a plethora of new applications -- ranging from dark spinors Hawking radiation across black strings and black holes, dark matter, cosmology, gravity on 5D  thick branes, particle physics and condensed matter \cite{lee1,daSilva:2012wp,exotic,Jardim:2014xla,Liu:2011nb}. 

On the other hand, Fierz identities are constraints on bilinear covariants, making it possible to classify spinors fields on manifolds of arbitrary dimensions and signatures.  Recently, a classification of spinor fields on  Lorentzian \cite{BR2016} and Riemannian 7-manifolds has been constructed, emulating and generalising the Lounesto classification, that solely holds in Minkowski spacetime \cite{BBR,BR2016}. 
These unprecedented classifications provide new classes of spinors and potentially new solutions in supergravity (SUGRA), regarding, in particular, its compactification that results 7-manifolds. Since the Fierz identities can be used to provide 
quite few realisations \cite{1}, we shall discuss here the last mod 8 possibility, by studying the occurrence of new 
spinor fields on Lorentzian 5D manifolds. In fact, since it plays a prominent role on brane-world models and fermions localisation  \cite{Zhang,Andrianov:2013vqa,Liu0907.0910}, new fermions in a bulk are here proposed.    In particular, C, P, and T symmetries in the bulk have been used to localise fermions on the brane  
     \cite{Casadio:2001fe}. 
     
     Motivated by the vast important applications of the new spinors in Lounesto classification in the last decade, out of the Dirac spinors paradigm, we shall derive in this paper new spinors in a 5D bulk.  The Clifford-algebraic spinor structure, in the geometric Fierz identities, sharply controls the existence of new classes of spinors in arbitrary  dimension and metric signature  \cite{1}. In fact, although  7D Riemannian spacetimes classically only admit Majorana spinors \cite{1,livro}, we have found two new non-trivial classes of hidden spinors, in the context of the AdS$_4\times S^7$ compatification of  SUGRA \cite{BR2016,BBR}, based on the spinor bilinear covariants. One of the authors also proved  that such new 7D spinors can, still, be identified to Killing spinors on 7D Kerr black hole horizons  \cite{BR2016}. 
Hence we aim to classify spinor fields on Lorentzian 5-manifolds, that in particular realise the  space AdS$_5$ as the bulk of brane-world models. From an existing class of spinors in those spaces, we can obtain more six spinor classes,  
whose representatives are new classes of spinors in the bulk. 

This paper is organised as follows: in Sect. II, the bilinear covariants are the stage to revisit the classification of spinor fields in Minkowski spacetime, 
according  to the Lounesto classification prescription, and the Fierz
aggregate and its related boomerang are defined. In Sect. III, the geometric Fierz identities are employed and, from the admissible
pairings between spinor fields, the number of classes in the spinor field
classification is constrained, by the geometric Fierz identities \cite{1}. We study the case of spinor fields on
Lorentzian 5-manifolds and conclude that 5D spinors pertain to solely one class, through a similar approach to the one previously acquired in 7D \cite{BR2016,BBR,1}. 
 Spinor fields  on 5-manifolds can be then classified in eight classes, from the classification of   quaternionic spinor fields. One of the new classes encompasses the standard 5D spinor fields, and six others non-trivial classes provide new candidates  for physical solutions, for instance, in thick brane models and gravity on AdS$_5$. Finally, in Sect. IV, in the context of 5D Kerr black holes, the current density connects the null Killing vector at the horizon to the time Killing vector, at spatial infinity.  In Sect. V we present our concluding remarks, discussion, and perspectives. 

\section{Bilinear Covariants in Minkowski Spacetime}

In order to fix the notation, let $(M,g)$ be an oriented spin manifold, equipped with a metric $g$ of signature $p-q$. On the exterior bundle $\Omega(M)=\oplus_{i=0}^\infty\Omega_i(M)$ associated to 
the manifold, standard endomorphisms are defined, for an arbitrary homogeneous $k$-form  $a \in \Omega^k(M)$: a) the reversion, $\tilde{a}=(-1)^{[[k/2]]}a,$ where $[[k]]$ is the integer part of $k$; b) the grade 
involution, $\hat{a}=(-1)^{k}a$; and c) the exterior algebra conjugation, defined by the composition of the previous morphisms, namely, $\bar{a}\equiv\tilde{\hat{a}}$. These anti-automorphisms and automorphism can be forthwith extended, by linearity,  
to the whole exterior bundle. Spinor fields and their bilinear covariants are fundamentally constructed in  the Clifford bundle framework. Firstly, $\Omega(M)$ must be endowed with the Clifford product 
$u \circ a = u \wedge a+ u 
\lrcorner a$, for all form fields $u \in \Omega^1(M)$ in the cotangent bundle,  where $\lrcorner$ is the left contraction ${g}(a \lrcorner b,c):={g}(b
,\tilde{a}\wedge c)$, where $g$  here also denotes  the extension of the metric on the bundle $\Omega(M)$, and $b ,c \in  \Omega(M)$ are arbitrary form fields. The Hodge operator $\star:\Omega(M)
\rightarrow\Omega(M)$ shall be used throughout the paper, defined by  
$(\star a)\w  b = g(a,b)$ \cite{livro}.

The spinor bundle in four Lorentzian dimensions can be represented by the vector bundle $\mathcal{P}_{\mathrm{Spin}_{1,3}^{e}}(M)\times
_{\upsigma}\mathbb{C}^{4}$. Classical spinor fields $\psi \in \sec \mathcal{P}_{\mathrm{Spin}_{1,3}^{e}}(M)\times_{\upsigma
}\mathbb{C}^{4}$ carry the
${\left(\frac12,0\right)}\oplus{\left(0,\frac12\right)}$ irreducible representation $\upsigma$ of the usual Lorentz group.  The bilinear covariants are sections of the exterior bundle $\Omega(M)$, {whose basis 
is given by $\{\theta^\alpha\}$}  \cite{lou2} (hereupon $\alpha,\beta = 0,1,2,3$):
\begin{subequations}
\begin{eqnarray}
\sigma &=& \bar{\psi}\psi\in\sec\Omega^0(M)\,,\label{sigma}\\
\mathbf{J}&=&{{\rm J}_{\a }\theta ^{\a }=\bar{\psi}\gamma _{\a }\psi\, \theta
^{\a}}\in\sec\Omega^1(M)\,,\label{J}\\
\mathbf{S}&=&S_{\a \b }\theta ^{\a}\wedge\theta^{ \b }=\tfrac{1}{2}i\bar{\psi}\gamma _{\a
\b }\psi \,\theta ^{\a }\wedge \theta ^{\b }\in\sec\Omega^2(M)\,,\label{S}\\
\mathbf{K}&=& K_{\a }\theta ^{\a }=i\bar{\psi}\gamma_{5}\gamma _{\a }\psi
\,\theta ^{\a }\in\sec\Omega^3(M)\,,\label{K}\\\omega&=&-\bar{\psi}\gamma_{5}\psi\in\sec\Omega^4(M)\,,  \label{fierz.}
\end{eqnarray}\end{subequations}
where $\gamma_5:=i\gamma_0\gamma_1\gamma_2\gamma_3$ denotes the Minkowski spacetime volume element, and  the spinor conjugation reads  $\bar\psi=\psi^\dagger\gamma_0$ \cite{livro}. Physical observables, exclusively for the Dirac electron theory, are realised by bilinear covariants. The gamma matrices  satisfy the Clifford relation  $\gamma_{\a }\gamma _{\b
}+\gamma _{\b }\gamma_{\a }=2g_{\a \b }\mathbf{1}$, {where $g_{\a\b}$ denotes the 
Minkowski spacetime metric components.}
 In fact, the current density $\mathbf{J}$, the spin density $\mathbf{S}$, and the chiral current $\mathbf{K}$ satisfy the Fierz identities \cite{lou2} 
\begin{equation}\label{fifi}
-(\omega+\sigma\gamma_{5})\mathbf{S}=\mathbf{J}\wedge\mathbf{K},\qquad\mathbf{K}\lrcorner \mathbf{K}+\mathbf{J}\lrcorner\,\mathbf{J}
=0=\mathbf{J}\llcorner\mathbf{K},\qquad
\mathbf{J}\,\lrcorner\, \mathbf{J}=\omega^{2}+\sigma^{2}\,.  
\end{equation}
\noindent A spinor field is said to be singular if $\omega=0=\sigma$, and
regular, otherwise. Singular spinors encompass flag-dipole, flagpole and dipole spinors. The Dirac spinor is a type of regular spinor. Besides, in particular, Elko and Majorana spinor fields are particular realisations of flagpole spinors,  and Weyl spinors realise dipole spinors, however there are more flagpole, flag-dipole, and dipole spinor fields than those realisations. 
Majorana spinors (with canonical mass dimension 3/2) and Elko spinors (with mass dimension 1) are both type-5 spinor fields, in Lounesto classification of the spinor fields.  A bottom-up  approach, to find the lacking dynamics of the hidden spinors in Lounesto classes, has been to look for certain examples, like the flag-dipoles in ESK  models of gravity \cite{esk} that are solutions of the Dirac equation
in $f(R)$-torsional background. However, it is neither effective nor a systematic approach. Celebrated  procedures in Refs. \cite{Cavalcanti:2014wia,fabbri} consist of laudable efforts to point out a direction to derive the dynamics that rule each subclass of all spinor classes, in Lounesto spinor fields classification.  

Lounesto derived, from the bilinear covariants, a classification of spinor fields \cite{lou2}. The condition  
$\mathbf{J}\neq0$ is satisfied in all cases, since ${\bf J}$ represents the current density, in particular, 
for Dirac spinors \footnote{Nevertheless, this condition can be disregarded for three extra classes recently found, consisting of mass dimension one spinors in Minkowski spacetime \cite{EPJC}, that are conjectured to be ghost spinors.}:
\beq
{\rm\qquad \underline{Regular \ \ spinors}:}&\left\{
\begin{array}{cc}
1)\;\omega\neq0,\;\;\;  \sigma\neq0,\;\;\;\mathbf{K}\neq 0, \;\;\;\mathbf{S}\neq0 &\\
\label{Elko11}
2) \;\omega = 0,\;\;\;
\sigma\neq0,\label{dirac1}\;\;\;\mathbf{K}\neq 0, \;\;\;\mathbf{S}\neq0\\
3)\;\omega \neq0, \;\;\;\sigma= 0,\label{dirac21}
\;\;\;\mathbf{K}\neq 0\;\;\;\mathbf{S}\neq0\;\;
\end{array}
\right.&\\
{\rm\qquad \underline{Singular \ \ spinors}:} \;\boxed{\omega=\sigma=0}\;\;\text{and}\;\;&\left\{\begin{array}{cc}
&4) \;\;\mathbf{K}\neq 0, \;\;\;\mathbf{S}\neq0  \;\;\text{(flag-dipole spinors)}\quad\qquad\\
&\!\!\!\!\!\!5) \;\;\mathbf{K}=0,\;\;\; \mathbf{S}\neq0 \;\; \text{(flagpole spinors)}\quad\qquad\\
&\!\!\!6)\;\; \mathbf{K} \neq 0,\;\;\; \mathbf{S}=0
\;\;\;  \;\; \text{(dipole spinors)}\quad\qquad\end{array}\right.
\eeq
These spinors in 4D Minkowski space make us ready to classify, in the next section, spinor fields on Lorentzian 5-manifolds, following a similar process that was used to classify spinor fields on Euclidean \cite{BBR} and Lorentzian \cite{BR2016} 7-manifolds. Moreover, from this classification new classes of spinors in a 5D bulk shall be constructed.
 
 The Fierz identities (\ref{fifi}) does not hold for singular spinors. In fact, since singular spinors present $\sigma=0=\omega$, the so-called Fierz aggregate reads 
 \begin{eqnarray}
\mathbf{Z}= \gamma_{5}(\omega+\mathbf{K})+i\mathbf{S}+ \mathbf{J} +\sigma \,, \label{Z}\label{zsigma}
\end{eqnarray}
and the Fierz identities are, hence, replaced by the equations 
\begin{eqnarray}
&&
\mathbf{Z}\gamma_{\mu}\mathbf{Z}=4{\rm J}_{\mu}\mathbf{Z},\quad{\mathbf{Z}}^{2}=4\sigma \mathbf{Z},\quad -i\mathbf{Z}\star \tilde{\mathbf{Z}}=-4\omega \mathbf{Z}\quad \mathbf{Z}i\gamma_{\mu\nu}\mathbf{Z}=4S_{\mu\nu}\mathbf{Z},\quad \mathbf{Z}i\star\widetilde{(\gamma_{\mu}\mathbf{Z})} =4K_{\mu}\mathbf{Z}.
\end{eqnarray}
Clearly these expressions are led to the standard Fierz equations (\ref{fifi}), when $\psi$ is a regular spinor. 
 
{The spinor representations $\left(\f{1}{2},0\right)$ and $(0,\f{1}{2})$ of the Lorentz group 
regard left and right-handed Weyl spinors, respectively. Observe that $\left(0,\f{1}{2}\right)\oplus \left(\f{1}{2},0\right)$ represents  transformations on Dirac spinors.  
% and takes values in $SL(2,\mathbb{C})$, %where one is conjugated of the other. gives the spin ...........
The geometric underlying content of the Fierz aggregate can be encompassed by physical considerations, when the connected to the identity component of the spin group Spin$^e$(1,3)$\simeq$ SL$_+(2,\mathbb{C})$ is taken into account.
%, which acts on Dirac spinors, %space generated by the gamma matrices $M(4,\mathbb{C})$ %,that is generated by the gamma matrices,
 The irreducible representation of the Lorentz group can be split into the homogeneous parts of the aggregate 
 %, i.e., to each term of the decomposition we make to correspond a k-vector,% and the boomerang structure can be explored more 
\cite{Ticc}}.
\begin{equation}\label{ccccc}
{\begin{tabular}{cccccc}
$\left[\left(0,\f{1}{2}\right)\oplus\left(\f{1}{2},0\right)\right]\otimes \left[\left(0,\f{1}{2}\right)\oplus\left(\f{1}{2},0\right)\right] =$&$\underbrace{(0,0)}\oplus $&$\underbrace{\left(\f{1}{2},\f{1}{2}\right)}\oplus$ &$\underbrace{(1,0)\oplus(0,1)}\oplus$ &$\underbrace{\left(\f{1}{2},\f{1}{2}\right)}$&$\oplus\underbrace{(0,0)}$
\\
&$\sigma$&${\bf J}$&${\bf S}$&${\bf K}$&$\omega$\\
%component number\qquad\qquad\qquad 16=&1&4&3\;+\;3&4&1
\end{tabular}} \end{equation}\medbreak It is worth to mention that the spin density admits left and right 
splitting of the representation, as shown in Eq. (\ref{ccccc}). In addition,  
\noindent{
 the spinor representations are obtained from the fundamental spinor representations $\left(\f{1}{2},0\right)$ and $(0,\f{1}{2})$. 
For example, $(\f{1}{2},\f{1}{2})$ represents the set of boosts and rotations on the current density ${\bf J}$, that can be represented by $\psi^*_R\s^\m\psi_R$ and  $\psi^*_L\bar{\s}^\m\psi_L$, for left $(\psi_L)$ and right ($\psi_R$) spinors. The  $(1,0)\oplus (0,1)$ representation is related to anti-symmetric irreducible representations of the Lorentz group. 
 }
\section{New Classes of Spinor fields in Lorentzian 5-dimensional space}

The K\"ahler-Atiyah bundle ($\Omega(M),\circ$) on the manifold $(M,g)$ is employed to construct   the spin bundle $S$, where the module structure requires a morphism $\g:(\Omega(M),\circ)\rightarrow(\text{End}(S),\Ganz)$ \cite{1}. The bundle $S$ can be split into sub-bundles, $S_0$ and $S_1$, that have $\pm 1$ eigenvalues, respectively \cite{1,livro}.   
Local endomorphisms  $J_i$ denote quaternionic structures, $(i=1,2,3)$, on the bundle $S$, that commute with arbitrary exterior bundle elements, and satisfy the relations\footnote{The Einstein summation convention shall be used in the indexes $\{i,j,k\} = \{1,2,3\}$, hereupon.} \cite{1}
$$J_i\Ganz J_j=-\d_{ij}{\rm id}_S+ \e_{ij}^{\;\,k}J_j,$$
where $\e_{ijk}$ stands for  the Levi-Civita symbol. The representation of $\gamma$ is a graded involution $\g(\nu)=\pm{\rm id}_S$, where $\nu$ denotes the volume element on a 5-manifold \cite{1}.  Those complex structures $J_i$ provide the existence and {the admissibility of non-degenerate spinor bilinear forms \cite{1,alek}:}
\be \label{b1b2}
B_\m=B_0\,\Ganz\,({\rm id}_S\otimes J_\m)\;,
\ee where $J_0 ={\rm id}_S$ and $B_0=B$ { is the bilinear form that 
endows the 5D spin bundle. } {{A bilinear form $B$ is called admissible if a) $B$ it is either symmetric or skew-symmetric; b) $B$ induces  an extended reversion as an anti-automorphism of the K\"ahler-Atiyah algebra  
 \cite{1,alek,livro}. An additional item, in the definition of admissibility, holds for  the metric signature $p-q=0, 4, 6, 7\mod 8$: in these signatures, the spin bundle can be split into chiral real sub-bundles, namely, $S=S^+\oplus S^-$,  wherein $S^\pm$ are either $B$-isotropic or $B$-orthogonal with respect to each other, likewise  \cite{1,alek,livro}. However, these signatures are not going to be employed here.}}
 
Hence, a 5D analogue of Lounesto spinor fields classes provides, indeed, a different spinor fields classification. In fact, there exists, at least, two choices in Eq. (\ref{b1b2}), which are equivalent under the Hodge duality \cite{1}. 
We are concerned with a classification for quaternionic spinor fields, according to bilinear covariants on Lorentzian 5-manifolds.  
 
Let $A_{\psi\cdot\psi'}$ be an endomorphism on $S$, defined by $A_{\psi\cdot\psi'}(\psi''):=B(\psi'',\psi')\psi$. Then, in a general way, the geometric Fierz identities read  \cite{1}
$
A_{\psi_1\cdot\psi_2}\circ A_{\psi_3\cdot\psi_4}=B(\psi_3,\psi_2)A_{\psi_1\cdot\psi_4}
$, with an unique decomposition $ 
A_{\psi\cdot\psi'}=J_\m\Ganz A_{\psi\cdot\psi'}^{(\mu)}$ \cite{1}. In the bulk, the  components are expressed in terms of the bilinear covariants, up to a sign, as
\beq
A_{\psi\cdot\psi'}^{(0)}&=&\f{1}{8} B(\g_{I}\psi,\psi')\theta^{I},\\
A_{\psi\cdot\psi'}^{(i)}&=&\f{1}{8} B((\g_{I}\Ganz J_i)\psi,\psi')\theta^{I},
\eeq\noindent where hereon $I\in\{{\scriptstyle \emptyset,\mu,\mu\nu,\mu\nu\rho, \mu\nu\rho\sigma,  \mu\nu\rho\sigma\tau}\}$ is a composed index for a Lorentzian 5-dimensional manifold, for $\Gamma_{\emptyset}=\boldsymbol{1}$. Moreover, the notation { $\gamma_{\sigma_1\ldots
\sigma_j}= \gamma_{\sigma_1}\ldots
\gamma_{\sigma_j}$} and { $\theta^{\sigma_1\ldots
\sigma_j}= \theta^{\sigma_1}\wedge\cdots\wedge
\theta^{\sigma_j}$} is adopted. 
Hence,  spinor fields constraints  are evinced by the geometric Fierz identities \cite{1} 
\beq
g_{\mu\nu} \left(A_{\psi_1\cdot\psi_2}^{(\mu)}\circ A^{(\nu)}_{\psi_3\cdot\psi_4}\right)
%+A_{\psi_1\cdot\psi_2}^{(0)}\circ A^{(0)}_{\psi_3\cdot\psi_4}
&=&B(\psi_2,\psi_3)A_{\psi_1\cdot\psi_4}^{(0)} \\ 
\e^i_{\;jk}A_{\psi_1\cdot\psi_2}^{(j)}\circ A^{(k)}_{\psi_3\cdot\psi_4}+A_{\psi_1\cdot\psi_2}^{((0)}\circ A^{(i))}_{\psi_3\cdot\psi_4}&=&B(\psi_2,\psi_3)A_{\psi_1\cdot\psi_4}^{(i)}
\eeq
 {For the case of a 5D bulk,} the admissibility of the bilinear $B$ holds, if $B$ satisfies \cite{1}
\beq
B (\psi,\gamma_{\sigma_1\ldots
\sigma_j}\psi)= B (\gamma_{\sigma_1\ldots
\sigma_j}\psi,\psi)=(-1)^{\f{k(k-1)}{2}}B (\psi,\gamma_{\sigma_1\ldots
\sigma_j}\psi)
\eeq
 Therefore, bilinear covariants are constrained by the expressions  $\varphi_j:= \big\vert A^{(0),j}_{\psi \cdot\psi}\big\vert$, namely 
 \beq
\label{ddd}
\varphi_j= \frac{1}{j!}B (\psi,\gamma_{\sigma_1\ldots
\sigma_j}\psi){\;\theta^{\sigma_1}\wedge\cdots\wedge \theta^{\sigma_j}}, \eeq\noindent and $\varphi_j$ is equal to zero, unless if
$k(k-1)\equiv 0   \mod 4$. It means that $\varphi_j$ equals zero,  unless $k=0,1,4,5$. 
 
Three further bilinear mappings are constructed from the quaternionic structures $J_i$, through the commutativity $[J_i, \g_j]=0$, and the identity $J_i^\intercal=-J_i$,
 \beq
 B(\psi,J_i\Ganz\g_{I}\psi)&=&B(\g_{I}\Ganz J_i\psi,\psi)\\
 B(\g_{I}\Ganz J_i\psi,\psi)&=&(-1)^{\f{l(l-1)}{2}+1}B(\psi,J_i\Ganz\g_{I}\psi)
 ,\eeq
 where $l=|I|$. It implies that an additional quaternionic bilinear covariant, $\phi_l^i$, emulates the previous ones, being  defined by
 \be \label{ddd1}
 \phi_l^i= \frac{1}{l!}B (\psi,J^i\Ganz\gamma_{\sigma_1\ldots
\sigma_l}\psi)\;{\theta^{\sigma_1}\wedge\cdots\wedge \theta^{\sigma_l},}
\ee
vanishes, unless $l(l-1)+2\equiv 0  \mod 4$, namely, unless $l=2,3$. Hence, more six non-trivial   bilinear covariants, $\phi_2^i$ and $\phi_3^i$, are acquired, accordingly. In addition, the Hodge dual operator provides  the following dualities:
$$
\varphi_5=\star\varphi_0,\;\; \quad\varphi_4=\star\varphi_1,\;\;\quad\phi^i_3=-\star\phi^i_2. 
$$ Moreover, the following inhomogeneous forms \cite{1}
\be
\om_0=\f{1}{16}(\varphi_5+\varphi_4+\varphi_1+\varphi_0),\qquad\quad\om_i=\f{1}{16}(\phi^i_2+\phi^i_3)
\ee
can be used together with the Hodge duality, yielding the generators $\om_j$ to be reduced to $ 
\om_0=2\f{1}{2}(1+\star)(\om_0^{\mathring{}})$ and $\om_i=2\f{1}{2}(1+\star)(\om_i^{\mathring{}}),$  where $\om^{\mathring{}}_0=\f{1}{16}(\varphi_1+\varphi_0)$ and $\om^{\mathring{}}_i=\f{1}{16}\phi^i_2$ are the truncated generators of the truncated algebra $(\om^{\mathring{}}(M),\bullet)$ \cite{1}, which is defined from the algebra $(\om(M),\circ)$ by the homomorphism 
$
\f{1}{2}(1+\star)(\varphi\bullet\psi)=\f{1}{2}(1+\star)(\varphi)\circ\f{1}{2}(1+\star)(\psi).
$
The Fierz identities assume the form \cite{1}:
\begin{subequations}\beq\label{fierz}
\varphi_1\;\bullet\;\varphi_1-\delta_{ij}\phi^i_2\;\bullet\; \phi^j_2=7\varphi_0^2+6\varphi_0\varphi_1,\\\label{fierz1}
\vp_1\Db\phi_2^i+\phi_2^i\Db\vp_0+\e^{i}_{\;\,jk} \phi_2^j\Db\phi_2^k=6\vp_0\phi_2^i.
\eeq \noindent
\end{subequations}\noindent Those expressions can present an intuitive framework, by introducing  the product $\Delta _j:\Omega (M)\times \Omega
(M)\rightarrow \Omega (M)$ \cite{1}: 
\beq
(j+1){\zeta\,\,\Delta
_{j+1}\,\tau\,=g^{ab}(\gamma_{a}\lrcorner\zeta\,)\,
\Delta_j\,(\gamma_b\lrcorner\tau\,)\,,\qquad\zeta\,, \tau\,\in \Omega (M)\,,}
\eeq\noindent where $g^{ab}$ denotes the {5D} metric  tensor coefficients. 
By denoting $\Delta _0$ the exterior product,  the next terms
 of this iteration read
 \beq
{\zeta\,\,\Delta
_1\,\tau\,}&=&{g^{ab}(\gamma_{a}\llcorner\zeta\,)\w(\gamma_b\llcorner\tau\,)\,,\qquad\quad 
\zeta\,\,\Delta _2\,\tau\,=
\f{1}{2}g^{ab}g^{cd}[(\gamma_{a}\wedge\gamma_c)\llcorner\zeta\,]\w[(\gamma_b\wedge\gamma_d)\llcorner\tau\,)]\,}
.\eeq
Hence it yields \cite{1}:
\beq
\label{6}\delta_{ij}\big(\phi^i_2\Db\phi^j_2&-&\star(\phi^i_2 \wedge\phi^j_2)+(\phi^i_2)^*\phi^j_2\big)=0,\\
\label{7}\vp_1\Db\phi^i_2+\phi^i_2\Db\vp_1&=&2\star(\vp_1\wedge \phi^i_2),\\ 
\e_{ijk}\Big[\phi^j_2\Db\phi^k_2&-& \left(\star(\phi^j_2\wedge\phi_2^k-\phi_2^j\;\D_1\; \phi_2^k-\phi_2^j\;\D_2\; \phi_2^k\right)\Big]=0.
\eeq
The expression $\vp_1\lrcorner\phi^i_2=0$ yields $\star(\vp_1\wedge\phi_2^i)=a\;\phi_2^i$, implying that eqs. (\ref{fierz}, \ref{fierz1}) read:
\beq \label{1}
\vp_1^*\vp_1+\delta_{ij}\phi_2^i\phi_2^j&=&7\vp_0^2,\\\label{2} \delta_{ij}\star(\phi_2^i\wedge\phi_2^j)&=&-6\vp_0\vp_1\\
\label{3}2\star(\vp_1\wedge\phi_2^i)-\e^i_{\;jk}(\phi_2^j\,\Delta_1\,\phi_2^k)&=& 
6\vp_0\phi_2^i,\\\label{4}\e_{ijk}\star(\phi_2^j\wedge\phi_2^k)&=&0\,.\eeq

Some constraints to the forms $\vp_1$ and $\phi_2^i$ are obtained from the expression $\delta_{ij}\vp_1\Db\phi_2^i\Db\phi_2^j$. By substituting Eq. (\ref{2}) into Eq. (\ref{6}),  it yields 
\beq\label{phi12}
\delta_{ij}(\vp_1\Db\phi_2^i)\Db\phi_2^j=a\;\delta_{ij}\phi_2^i\Db\phi_2^j=-6a\;\vp_0\vp_1 -a\;\delta_{ij}\phi^i_2\phi^{j}_2.
\eeq
On the other hand, Eq. (\ref{2}) implies that
\beq\label{phi122}
\delta_{ij}\vp_1\Db(\phi_2^i\;\Db\;\phi_2^j)=-\delta_{ij}\phi^i_2\phi^j_2\vp_1-6\;\vp_0(\vp_1\,\Db\,\vp_1)
.\eeq
Comparing Eq. (\ref{phi12}) with Eq. (\ref{phi122}) yields
$
6a\,\vp_0=\delta_{ij}\phi^{*i}_2\phi^j_2=6\;\vp_0||\vp_1||^2/a$. Subsequently, it yields $-7\vp_0^2+6a\vp_0+a^2=0$, 
with solutions $a=\vp_0$ and $a=-7\vp_0$. The last one is not compatible \cite{1}, being  only the solution $a=\vp_0$  useful. Finally, the Fierz identities (\ref{fierz},\ref{fierz1}) yield  the following system of constraints to the forms $\vp_1$ and $\phi_2^i$ \cite{1}:
\beq
||\vp_1||^2&=&\frac{1}{36}\delta_{ij}\phi^i_2\circ\phi^j_2=\vp_0^2,\quad\;\;\;\;\;\;\;\;\;-2\vp_1\lrcorner\left(\f{1}{2}(1+\star)\phi^i_2\right)=4\vp_0\phi^i_2=\e^{i}_{\;jk}\phi_2^j\Db\phi_2^k,\\\!\!\!\!\!\!\!\!\!\!\!\!\!\!\!\!\!\!\!\!\!\!\!\!\!\delta_{ij}\phi_2^i\wedge\phi_2^j&=&-6\vp_0\star\vp_1\,.
\eeq
By normalising the spinor $\psi$, such that $\vp_0=B(\psi,\psi)=1$, the classification of spinors on 5D Lorentzian manifolds is then derived. More specifically, there exists only a single class of quaternionic spinor in the 5D Lorentzian Clifford bundle, reading: 
\beq
\!\!\!\!\!\!\!\!\!\!\!\!\!\!\varphi_0\neq 0, \quad \varphi_1\neq 0, \quad\varphi_2= 0, \quad\varphi_3= 0,\quad
\varphi_4\neq 0,\quad\varphi_5\neq 0, \label{class}\\
\!\!\!\!\!\!\!\!\!\!\!\!\!\!\phi_0 =0, \quad \phi_1= 0, \quad\phi_2\neq 0, \quad\phi_3\neq 0,\quad
\phi_4= 0,\quad\phi_5= 0,\label{class1}
\eeq
where $\phi_n\neq 0$ indicates that $\phi^k_n\neq 0$, for at least one value of $k$; and $\varphi_i,\phi_i^k\in\Omega^i(M).$ By the Hodge duality $\varphi_n= \star\varphi_{5-n}$, only half of bilinears is  necessary, 
\beq
\varphi_0\neq 0, \quad \varphi_1\neq 0, \quad\varphi_2= 0,\quad\phi_0^k =0, \quad \phi^k_1= 0 \text{\,  and } \quad\phi^k_2\neq 0
\eeq 
holding for at last one value of $k$. Complexification increases the number of spinor classes \cite{BBR}. The  components of the bilinears, for the quaternionic case, read \cite{1}:
\begin{subequations}
\beq
 B (\psi,\gamma_{\sigma_1\ldots\sigma_{2j}}\psi) 
&=&
\f{1}{2}\left[
 B\left(J_1\psi,\gamma_{\sigma_1\ldots\sigma_{2j}}\;\Ganz\;J_1\psi\right)-  B\left(\psi,\gamma_{\sigma_1
\ldots\sigma_{2j}}\psi\right)\right]\,,\\
 B (\psi,\gamma_{\sigma_1\ldots\sigma_{2j+1}}\psi) 
&=&\f{1}{2}\left[B\left(\psi,\gamma_{\sigma_1\ldots\sigma_{2j+1}}\;\Ganz\;J_2\psi\right)-  B\left(J_1\psi,\gamma_{\sigma_1
\ldots\sigma_{2j+1}}\;\Ganz\;J_3\psi\right)\right]\,,\\
 B(\psi,J_k\,\Ganz\,\gamma_{\sigma_1\ldots\sigma_{2j}}\psi) 
&=&\f{1}{2}\left[B\left(J_i\psi,\gamma_{\sigma_1\ldots\sigma_{2j}}\;\Ganz\;J_i\psi\right)-  B\left(\psi,\gamma_{\sigma_1
\ldots\sigma_{2j}}\psi\right)\right]\,, \\
B(\psi,J_k\,\Ganz\,\gamma_{\sigma_1\ldots\sigma_{2j+1}}\psi)&=&\f{1}{2}\left[B\left(J_i\psi,\gamma_{\sigma_1\ldots\sigma_{2j+1}}\;\Ganz\;J_k\psi\right)-  B\left(\psi,\gamma_{\sigma_1
\ldots\sigma_{2j+1}}\;\Ganz\;J_l\psi\right)\right]\,.
\label{fgh}
 \eeq
 \end{subequations}
 The higher degree 
generalisation of $B$ is provided, according to \citep{1}, by: 
 \beq\label{formac}
\upbeta_j(\psi,\gamma_{\sigma_1\ldots\sigma_j}\psi')= B\left(\psi,\gamma_{\sigma_1\ldots\sigma_j}\;\Ganz\;J_1\psi'\right)-
iB\left(\psi,\gamma_{\sigma_1\ldots\sigma_j}\;\Ganz\;J_3\psi'\right)\,.
\eeq\noindent 
The bilinear covariants can be extended, by identifying
\beq
\upvarphi_j&:=&
\frac{1}{j!}\upbeta_j(\psi,\gamma_{\sigma_1\ldots\sigma_j}\psi)\theta^{
\sigma_1}\wedge\cdots\wedge\theta^{\sigma_j}\,\\
 \upphi_l^i&:=& {\frac{1}{l!}\upbeta_l \left(\psi,J^i\Ganz\gamma_{\sigma_1\ldots
\sigma_l}\psi\right)\theta^{\sigma_1}\wedge\cdots \wedge\theta^{\sigma_l}}\,.
\eeq 
%Since both terms in the real part and also both
%terms in the imaginary part in (\ref{formac}) %as well
% can cancel each other, in the complex version it is possible that the bilinears $\varphi_0,$  $\varphi_1$ and $\phi_2$ be equal to zero. 
Hence, eight classes of spinor fields $\psi\in S$  
 are found, however one class regards the trivial spinor.  Concerning the
classification of spinor fields on Lorentzian 5-manifolds, the
 above derived constraints  are used to evince the Lorentzian 5D classification:
\begin{subequations}
\beq
\upvarphi_0\neq 0, \quad \upvarphi_1\neq 0, \quad\upphi^k_2\neq 0,\label{c11}\\
\upvarphi_0\neq 0, \quad \upvarphi_1\neq 0, \quad\upphi^k_2=0,\\
\upvarphi_0\neq 0, \quad \upvarphi_1= 0, \quad\upphi^k_2\neq 0,\\
\upvarphi_0\neq 0, \quad \upvarphi_1= 0, \quad\upphi^k_2=0,\\
\upvarphi_0= 0, \quad \upvarphi_1\neq 0, \quad\upphi^k_2\neq 0,\\
\upvarphi_0= 0, \quad \upvarphi_1\neq 0, \quad\upphi^k_2= 0,\\
\upvarphi_0= 0, \quad \upvarphi_1= 0, \quad\upphi^k_2\neq 0,\\
\upvarphi_0= 0, \quad \upvarphi_1= 0, \quad\upphi^k_2=0.\label{c18}
\eeq
\end{subequations} It is implicit in (\ref{c11}-\ref{c18}) that {$\{\upvarphi_l,\upphi_l\}=\{0\}$}, for $l=1,2,3,5$. 

In the next section, an application of these new spinor fields shall be derived. The current density, that is the vector bilinear covariant, is written in terms of components of spinors and can be identified as a Killing spinor field on a black hole horizon.
\section{Spinors on (1,4)-dimensional Kerr spacetime}
 
 Regarding Kerr black holes, the null Killing vector on the event horizon reaches the time Killing vector at the spatial infinity through a specific vector field $\xi^\m$ \cite{Mei2}. For this reason, it is supposed the existence of a 1-form current density that is preserved, i.e., the spinor $\psi$ satisfies the Dirac equation, however it is not necessary $\psi$ to be a Dirac fermion \cite{Mei2}. In this section, % some resultsand, then,
we find a expression to the time Killing vector in terms of the spinor components in its \emph{f\"unfbein} basis $e^A={e^A}_\m dx^\m$ for a 5D spacetime. The covector field $\xi^\m$ is constructed from the current density ${\rm J}^\m$ for the spinor field $\psi$%, that satisfies the Dirac equation,
\be 
\xi^\m=b_\psi {\rm J}_\psi^\m= b_\psi \bar{\psi} \g^\m\psi,\qquad\quad \g^\m={e_{A}}^\m \g^A,
\ee
where $b_\psi$ is a constant and $\g^A$ denotes the $(1,4)$-dimensional gamma matrices in the \emph{f\"unfbein} basis, represented as \cite{Mei2}
\be\label{gamma} 
\g^0=\s^1\otimes {\bf 1}_2,\quad \g^k=i\s^2\otimes \s^k,\quad \g^4=i\s^3\otimes{\bf 1}_2. 
\ee
If the spinor field $\psi$ satisfies
{$
(\g^0\pm \g^6)\psi=0
,$ where $\g^6:=-\g^1\g^2\g^3\g^4\g^0$,}
then the \emph{f\"unfbein} basis yields $
\psi=(\a, \a)^\intercal$,  %=(1,1 )^\intercal\otimes \a,$
 where ${\a}=(\a_1,\a_2)^\intercal$ and $\a_1,\a_2$ are scalars \cite{Mei2}.   
The most general metric for a 5D axisymmetric ,  stationary, black hole assumes the following form \cite{Mei2}
\beq 
ds^2\!=\!-h_t(dt\!+\!h_1d\phi_1\!+\!h_2d\phi_2)^2\!+\!h_rdr^2\!+\!h_\theta d\theta^2\!+\!g_{22}d_{\omega_2,\phi_2}^2
\!+\!g_{11}\left(d_{\omega_1,\phi_1}\!+\!g_{12}d_{\omega_2,\phi_2}\right)^2,
\eeq
where $d_{\omega_a,\phi_a}:=\omega_adt-d\phi_a$ ($a=1,2$), and the metric coefficients 
only depend on $r$ and $\theta$; as usual one denotes the time coordinate by $t$, and the radius $r$, the latitudinal angle $\theta$, and
the azimuthal angles $\phi_1$ and $\phi_2$. In this context, $\omega_1$ [$\omega_2$] is the angular velocity in the direction $\phi_1$ [$\phi_2$]. The metric, in terms of \emph{f\"unfbein}, reads \cite{Mei2}
$
ds^2=\eta_{AB}e^Ae^B,$ for $A,B=0,\ldots, 4
,$ 
where the \emph{f\"unfbein} $e^A={e^A}_\m dx^\m$ is given by  
\beq
e^0=h_t^{1/2}(dt+h_1d\phi_1+h_2d\phi_2), \quad e^1=h_r^{1/2}dr,\quad e^2=h_\theta^{1/2}d\theta,\\
e^3=g_{11}^{1/2} 
\left[d_{\omega_1,\phi_1}-g_{12}d_{\omega_2,\phi_2}\right], \quad e^4=-g_{22}^{1/2}d_{\omega_2,\phi_2}.\eeq
%That expression can be written in terms 
Explicitly, 
%The components of the corresponding 
\be\label{e}
[{e^A}_\m]=\left(\begin{tabular}{ccccc} 
$h_t^{1/2}$&0&0&$h_t^{1/2}h_1$&$h_t^{1/2}h_2$\\
0&$h_r^{1/2}$&0&0&0\\
0&0&$h_\theta^{1/2}$&0&0\\
$-g_{11}^{1/2}(\omega_1+g_{12}\omega_2)$&0&0&$g_{11}^{1/2}$&$g_{11}^{1/2}g_{12}$\\
$-g_{22}^{1/2}\omega_2$&0&0&0&$g_{22}^{1/2}$
\end{tabular}\right). 
\ee 
%where we choose $(\p_0,\ldots,\p_4)=(\p_t,\p_r,\p_\theta,\p_{\phi_1},\p_{\phi_2}).$ 
Its inverse matrix $[{e^A}_\m]^{-1}$ %of (\ref{e}) 
is used to get the components of the required vector field $\xi^\m$. % are scalar for each value $A$, then 
By denoting the column matrix $[\bar{\psi}e^A\psi]$, then
\be
\xi^\m= [{({e^{-1})}^\m}_{A}][\bar{\psi}\theta^A\psi]=\left(1,0,0,\omega_1,\omega_2\right)^\intercal= \p_t+\omega_1\p_{\phi_1}+\omega_2\p_{\phi_2}.\ee
 Finally, the current density vector field can also be expressed in the \emph{f\"unfbein} components, reading
\be
{\rm J}^\m\p_\m=(\bar{\psi}\g^\m\psi)\p_\m=2(|\a_1|^2+|\a_2|^2)\p_t,
\ee
where the gamma matrices  are in the representation (\ref{gamma}). %and the reduction $\psi=(1,1)^\intercal \otimes \a$ is adopted in the computations.
 We can compare this result to a similar one in Ref. \cite{BR2016},  where the current density acquires a similar expression  on Lorentzian 7-manifolds.

{Regarding this classification of spinor fields on Lorentzian 5-manifolds, there exists only one non-zero bilinear covariant with quaternionic structure, namely $\phi_2^k$ in Eq.(\ref{ddd1}), that reads 
\begin{equation}\phi_2^k=\f{i}{2}\left(g_{j\mu}g_{\nu}^{\;k}-g^k_{\;\mu}g_{\nu j}\right)\bar\psi(\sigma^j\otimes 1_2)\psi\;\theta^\m\wedge\theta^\n
,\end{equation}}
with respect to the \emph{f\"unfbein} and the realisation $J_i=i\s_i\otimes 1_2$ of the complex structure, where $\s_i$ are the Pauli matrices, 
{ that results $\psi=(\a,\a)^\intercal$. It implies that the bilinear covariants $\phi_2$ are}
\begin{subequations}
\beq
\phi^1_2&=&-1_2\otimes\left[\left(\a^\dagger_1\a_2+\a^\dagger_2\a_1\right)\gamma_{01}+i\left(\a^\dagger_2\a_1-\a^\dagger_1\a_2\right)\gamma_{02}+\left(|\a_1|^2-|\a_2|^2\right)\gamma_{03}\right]\\
\phi^2_2&=&-i1_2\otimes\left[\left(\a^\dagger_1\a_2+\a^\dagger_2\a_1\right)\gamma_{01}+i\left(\a^\dagger_2\a_1-\a^\dagger_1\a_2\right)\gamma_{02}+\left(|\a_1|^2-|\a_2|^2\right)\gamma_{03}\right]=i\phi^1_2\\
\phi^3_2&=&-i1_2\otimes\left[\left(\a^\dagger_1\a_2+\a^\dagger_2\a_1\right)\gamma_{23}+i\left(\a^\dagger_2\a_1-\a^\dagger_1\a_2\right)\gamma_{31}+\left(|\a_1|^2-|\a_2|^2\right)\gamma_{12}\right]=i(\star\phi^1_2).
\eeq 
\end{subequations}
Therefore, $\phi^1_2, \phi^2_2$ and $\phi^3_2$ are equivalent.

These new spinors are coefficients of quantum fields that may play an important role to 
explore fermion localisation on brane-world models. In fact, except for 
two exceptions, all the literature regards bulk  spinors, generally in an AdS$_5$, or asymptotically AdS$_5$, bulk. Bulk spinors are then governed by the Dirac equations and here we studied a departure out of this paradigm. To explore localisation issues is beyond the scope of this paper, and shall be more discussed in the forthcoming section.

\section{Concluding remarks and outlook}

A more profound comprehension about spinor fields on Lorentzian 5-manifolds is potentially useful in brane-world models, through a classification of 5D  spinor fields.
They appear in the bulk and some classes of %possible
spinor fields are obtained, whose respective bilinear covariants current density 1-form $\varphi_1$ and the dual spin density 2-form with a quaternionic structure $\phi_2^i$ %give the classification and 
are computed, with respect a \emph{f\"unfbein} basis. 
The current density field is shown to be a null Killing vector at Kerr black hole horizon that reaches the spatial infinity as a time Killing vector. 

 The classification of spinors according to bilinear covariants on Minkowski spacetime, that is revised in Sect.  II, has opened a very useful way to discover unexpected 
new physical features and to propose new candidates for fermionic fields in Minkowski spacetime.  In Sect. III, we constructed bilinear covariants for spinors that are constructed on Lorentzian 5-manifolds. 
%This is important to study the important case of fermionic fields on Lorentzian 5-manifolds $AdS_5$. 
We exhibited a classification to that fields, following the same criteria already established on Minkowski 4D spacetime \cite{lou2}, on Euclidean 7-manifolds \cite{BBR}, and on Lorentzian 7-manifolds \cite{BR2016}. % We analyse the important case of spinor fields on . 
From that classification, we obtained six new other non-trivial classes, that can be models for new fermionic fields on 5-manifolds. Finally, null Killing 1-forms on  
5D Kerr spacetimes were analysed, and compared with previous incipient results in Ref.   \cite{Mei2}. The Killing vector is shown to be proportional to the current density. We write the current density 1-form $\varphi_1$ and a kind of spin 2-form with a quaternionic structure $\phi_2^i$
in terms of components in the \emph{f\"unfbein} basis. It is interesting to note also that their components exhaust all possible invariants that can be constructed with the studied  spinor fields.
% $\psi=(\a, \a)^\intercal$
%\be
%|\a_1|^2\pm|\a_2|^2\text{\;\; and\;\; }\a^\dagger_1\a_2\pm\a^\dagger_2\a_1.
%\ee
 %and we show that they are conserved quantities.
%{Refs. Unify, put in order}

Symmetry operators of the Dirac equation in curved backgrounds can be constructed from
Killing-Yano forms, that are  generalisations of Killing vector fields, which play an
important role in SUGRA. Hence, finding the solutions of the Killing
spinor equation gives way to determine the supergravity Killing spinors in relevant supergravity backgrounds \cite{Gauntlett:2007ts,Henningson:1998cd}. One
way to find the solutions of Killing spinor equation is constructing the symmetry operators -- solutions of the Dirac equation. Hence, regarding a  spinor field in, for example, AdS space, and the corresponding 
Dirac action in bulk, we can study solutions that dock new solutions of dynamical equations, that describe new spinors. 
A more complete analysis of fields in SUGRA AdS$_5\times S^5$ is possible, if spinor fields are classified also on Riemannian 5-manifolds. A further question that arises from our work concerns the nature of the bilinear $\phi_2^i$, since it emulates the (quaternionic) pseudospin of particles described by the spinors. A final answer can not be comprised upon the study of the respective equations of motion, that is beyond our scope here.  
\acknowledgments

KPSB acknowledges the UFABC and CAPES grants; RdR thanks FAPESP (grant No. 2015/10270-0) and CNPq (grant No. 303293/2015-2; No. 473326/2013-2) for partial support.

\end{document}